\documentclass[preprint,showpacs,showkeys,preprintnumbers,amsmath,amssymb]{revtex4}

\usepackage{graphicx}% Include figure files
\usepackage{dcolumn}% Align table columns on decimal point
\usepackage{bm}% bold math

\begin{document}

\title{Chaotic properties of the truncated elliptical billiard}
% Force line breaks  with \\

\author{V. Lopac}
\email{vlopac@marie.fkit.hr}
\author{A. \v Simi\' c}

\affiliation{Department of Physics, Faculty of Chemical Engineering and
 Technology,\\ University of Zagreb, Croatia%
}%

	     
\begin{abstract}
Chaotic properties of symmetrical two-dimensional stadium-like billiards
with elliptical arcs are studied numerically and analytically.  For the
two-parameter truncated elliptical billiard the existence and linear
stability of several  lowest-order periodic orbits are investigated in
the full parameter space. Poincar\' e plots are computed and used for
evaluation of the degree of chaoticity with the box-counting method. The
limit of the fully chaotic behavior  is identified with circular arcs.
Above this limit, for flattened elliptical arcs, mixed dynamics with
numerous stable elliptic islands is present, similarly as in the
elliptical stadium billiards. Below this limit the full chaos extends
over the whole region of elongated shapes and the existing orbits are
either unstable or neutral. This is conspicuously different from the
behavior in the elliptical stadium billiards, where the chaotic region
is  strictly bounded from both sides. To examine the mechanism of this
difference, a generalization to a novel three-parameter family of
boundary shapes is proposed and suggested for further evaluation.
\end{abstract}


\maketitle

\section{Introduction}

Two-dimensional planar billiards are nonlinear systems with rich and
interesting dynamical properties. A point particle, moving with constant
velocity within a closed boundary and exhibiting specular reflections on
the walls, can have regular, mixed or fully chaotic dynamics, in strong
dependence on details of the boundary shape. In physics, two-dimensional
billiards offer good examples of coexistence of regular, mixed and
chaotic dynamics in Hamiltonian systems. This type of behavior,
illustrated by the standard map and explained by means of the
KAM-theorem, is present in many realistic phenomena, such as planetary 
systems and various types of coupled oscillators\cite{Hamilt}. Chaotic
billiards were first introduced by  Sinai\cite{Sinai} who considered the
defocusing effects of circular scatterers in the two-dimensional Lorentz
gas.  After the important discovery by Bunimovich\cite{Bunist1,Bunist2}
that also the focusing  circular arcs can lead to a fully chaotic
behavior, many investigations were devoted to billiards with circular
arcs and, in a smaller extent, to other types of curved boundaries. The 
systematic mathematical description of chaotic billiards and an
extended  list of references can be found in the book by Chernov and
Markarian\cite{ChMar}. Rigorous investigations were concentrating on
methods for producing fully chaotic billiards and on specific properties
(Bernoulli, K-property, mixing and hyperbolicity) expressing differences
between chaotic systems\cite{ChMar,DelMaMar,Donnay,Mark,Woj}.  Various
aspects of billiard dynamics have been extensively examined during  last
decades\cite{Strel,Baecker,BeRo,Berry,DullRW,Heller,Makino,Prosen,
Reichl,Robnik}. In recent years,
properties  of classical billiards and their quantum-mechanical
counterparts  were used to explain and  improve performances of devices
in microelectronics and nanotechnology, especially of  optical
microresonators in dielectrical and  polymer
lasers\cite{Gmachl,Haray,Hentschel,Lebent,Stone,Tanaka}. 

We are stressing the fact that notable regions of full chaos have been 
discovered in billiards with elliptical arcs and piecewise flat
boundaries,  indicating that such billiards deserve further 
attention\cite{CanMarOKS,MarOKS,OKS,DMTEB}. In our previous work we
analyzed several types of billiards with noncircular arcs (parabolic,
hyperbolic, elliptical and generalized power-law), exhibiting mixed
dynamics\cite{LMRpar,LMRelh,LMRparlim}. Next we investigated, in the
full parameter space\cite{LPESB}, the elliptical stadium billiards
(ESB), first introduced by Donnay\cite{Donnay}. Here we extend  the same
type of analysis to the truncated elliptical billiards (TEB), which
although similar in appearance,  have different dynamical properties.
The truncated elliptical billiard (TEB)  is defined by a two-parameter
planar domain constructed by truncating an ellipse on  opposite sides
(Fig. 1). A symmetrical stadium-like shape thus obtained consists of a
rectangle with two elliptical arcs added at its opposite ends. The
corresponding billiard has been introduced by Del Magno\cite{DMTEB} who,
investigating a restricted part of the parameter space and applying the
mathematical method of invariant cones, determined the region of
hyperbolic behavior and  presented an estimate of the region where such
billiard could be ergodic. 

In the present paper we investigate numerically and analitically  the
truncated elliptical billiard (TEB) in the full parameter space, by
using two shape parameters $\delta$ and $\gamma$. This description of 
the billiard geometry and dynamics is consistent with our  previous
analysis  of the elliptical stadium billiard
(ESB)\cite{LMRpar,LPESB,LMRg1}, which is a two-parameter  generalization
of the Bunimovich stadium billiard\cite{Bunist2} and is a special case
of the mushroom billiard\cite{Mush2,BuniMush,Porter}. It has been
confirmed by analysis and numerical
computation\cite{Donnay,CanMarOKS,MarOKS,OKS,LPESB,LMRg1} that this
billiard is fully chaotic (ergodic) for a sizeable but strictly limited
region  in the parameter space,  defined by the stable two-bounce
horizontal periodic orbit on one and the pantografic orbits on the other
side. Our investigations of the ESB and TEB billiards confirm the
suggestion by Del Magno\cite{DMTEB} that in spite of apparently  similar
stadium-like shapes, these two billiards have essentially different 
dynamical properties. In the present  paper we describe our analytical
and numerical investigation of the truncated elliptical billiard  and
compare the obtained results with those for the elliptical  stadium
billiard.

In Section II we define the TEB billiard boundary and describe its
geometrical properties. In Section III the existence and stability of
selected orbits are discussed and illustrated by Poincar\' e plots and
orbit diagrams. In Section IV the Poincar\' e sections are used to
estimate, by means of the box-counting numerical method, the degree of
chaoticity for a given boundary shape. The results are shown in the
parameter-space diagram and compared with the same type of diagram for
the elliptical stadium billiard. In Section V we briefly discuss the
possible generalization of the truncated elliptical billiard providing a
transition between two types of the stadium-like elliptical billiards.
Finally, in Section VI we summarize the obtained results and propose
further investigations.

\section{Geometrical properties of the truncated elliptical billiard}

In our parametrization the truncated elliptical billiard (TEB) is
defined in the $x-y$ plane by means of the two parameters $\delta$ and
$\gamma$, satisfying   conditions $0\le\delta\le1$ and
$0<\gamma<\infty$. The billiard  boundary  is described as

\begin{equation} 
y(x)=\left\{ \begin{array}{ll} \pm  \gamma &  {   \rm  ,    if\    } 
 0\le|x|<\delta \\ 
 \\
\pm\gamma\sqrt{\frac{1-x^2}{1-\delta^2}} &
 {   \rm   ,    if\    }  \delta \le|x|\le1 \end{array} \right.
\end{equation}

\noindent The horizontal diameter is normalized to 2, so that the
horizontal semiaxis of the ellipse is 1. The vertical semiaxis of the
ellipse is $\gamma/\sqrt{1-\delta^2}$,  and the possible height  
$2\gamma$ of the billiard extends from $0$ to $\infty$. The horizontal
length of the central rectangle  is $2\delta$.

In special cases, for $\delta=0$ the shape is a full ellipse, for
$\delta=1$ it is rectangular, for $\delta=\gamma=1$ it is a square and
for $\delta=0$ and $\gamma=1$ a full circle. For
$\delta=\sqrt{1-\gamma^2}$ one obtains a set of truncated circle
billiards, which separates two distinct billiard classes, one for
$\delta<\sqrt{1-\gamma^2}$ with elongated elliptical arcs and the other
with $\delta>\sqrt{1-\gamma^2}$ and flattened elliptical arcs. Fig. 1
shows three typical shapes of the truncated elliptical billiard with
circular, flattened and elongated  elliptical arcs.

\begin{figure}
\caption{\label{fig:Fig1}Three types of the TEB billiard shape: (a)
circular shape, with $\delta=\sqrt{1-\gamma^2}$: $\delta=0.6$,
$\gamma=0.8$; (b) flattened shape  with elliptical arcs and
$\delta>\sqrt{1-\gamma^2}$: $\delta=0.6$,  $\gamma=2.2$ and (c)
elongated shape with $\delta<\sqrt{1-\gamma^2}$: $\delta=0.6$, 
$\gamma=0.2$ }

\end{figure}

The coordinates of the focal points are

\begin{equation} 
F\left[\pm\sqrt{\frac{1-\delta^2-\gamma^2}{1-\delta^2}},0\right] 
\end{equation}

\noindent for $\delta<\sqrt{1-\gamma^2}$, and

\begin{equation}
F\left[0,\pm\sqrt{\frac{\gamma^2+\delta^2-1}{1-\delta^2}}\right]
\end{equation}

\noindent for $\delta>\sqrt{1-\gamma^2}$.  They contain the important 
term $\tau=\gamma^2+\delta^2-1$ which is negative for
$\delta<\sqrt{1-\gamma^2}$,  positive for $\delta>\sqrt{1-\gamma^2}$,
and zero for $\delta=\sqrt{1-\gamma^2}$ (circular arcs). This limit is
shown  as the thick circular line in Fig. 2 presenting the structure  of
the $\gamma-\delta$ parameter space.

\begin{figure}
\caption{\label{fig:Fig2} Diagram of the two-dimensional parameter
space ($\gamma$, $\delta$). Lines denote the limits of existence and
stability for certain orbits, as explained in the text. }
\end{figure}

For $\delta<|x|\le1$ the  curvature radius is 

\begin{equation} 
R=\frac{[(1-\delta^2-\gamma^2)(1-x^2)+\gamma^2]^{3/2}}
{\gamma(1-\delta^2)}
\end{equation}

\noindent For $0\le|x|<\delta$ the boundary is flat and the curvature
radius is $R=\infty$, but for $|x|=\delta$ has a discontinuity and drops
to

\begin{equation}
R_{\delta}=\frac{[(1-\delta^2)^2+\gamma^2\delta^2]^{3/2}}
{\gamma(1-\delta^2)}
\end{equation}

\noindent At the endpoints of the horizontal axis of the ellipse
($|x|=1$)  the curvature radius is  

\begin{equation}
R_1=\frac{\gamma^2}{1-\delta^2}
\end{equation} 

\noindent which reduces to $R_1=1$ for circular arcs. For  full ellipses
with $\delta=0$ the curvature radius at $x=0$ is $R_0=1/\gamma$.  

As explained in \cite{LPESB}, the symbols $\theta$, $\phi$ and $\phi'$,
respectively, denote the angles which the normal, the incoming path  and
the outcoming path make  with the x-axis. The angle between the incoming
(or outcoming) path and the normal to the boundary is
$\beta=(\phi'-\phi)/2$. The angle between the tangent to the boundary at
the point T$(x,y)$ of impact and the incoming (or outcoming) path,
needed in the computation of the orbit stability, is 
$\alpha=(\pi/2)-\beta$.

The angles $\theta$, $\phi$ and $\phi'$  are connected by the relation

\begin{equation} 
\frac{2\tan\theta}{1-\tan^2\theta}=\frac{\tan\phi+\tan\phi'}
{1-\tan\phi\tan\phi'} 
\end{equation}

\noindent  The expression (7) is the basis for finding the existence
criteria for particular periodic orbits\cite{LPESB}. In our further
description we refer to the impact points T$(x,y)$ in the first
quadrant, with no loss of generality for the  obtained results. In the
Poincar\' e sections the points P$(x,v_x)$ are obtained by plotting the
slope of the velocity direction $v_x=\cos\phi$ versus  the x-coordinate 
of the intersection  point with the x-axis, as explained in
\cite{LMRparlim,LMRelh,LMRpar,LPESB}. The Poincar\' e diagrams obtained
in this way are area preserving.

As described in \cite{Berry,LPESB}, the stability of a periodic orbit is
assured if the absolute value of the  trace of the stability matrix $M$ 
is smaller than 2, thus if

\begin{equation}
-2<{\rm Tr} M<2
\end{equation}

\noindent Such orbits are elliptic, and those with $|{\rm Tr} M|=2$ are
neutral (parabolic). The stability matrix of the closed  orbit of period
$N$ can be written as  $M=M_{12}M_{23}..M_{N1}$, where the  $2\times2$
matrix $M_{ik}$ for  two subsequent impact points T$_i$ and T$_k$,
connected by a rectilinear chord of the length  $\rho_{ik}$, is

\begin{equation}
%\label{eq:}
M_{ik}=
\left(
\begin{array}{ll}
-\frac{\sin\alpha_i}{\sin\alpha_k}+\frac{\rho_{ik}}{R_i\sin\alpha_k}
 & -\frac{\rho_{ik}}{\sin\alpha_i\sin\alpha_k}\\
 & \\
-\frac{\rho_{ik}}{R_iR_k}+\frac{\sin\alpha_k}{R_i}+\frac{\sin\alpha_i}
{R_k} &
-\frac{\sin\alpha_k}{\sin\alpha_i}+\frac{\rho_{ik}}{R_k\sin\alpha_i}
\end{array}\right)
\end{equation}

\section{Classical dynamics of the truncated elliptical billiard}

\subsection{Billiards with $\delta<\sqrt{1-\gamma^2}$}

This subfamily of truncated elliptical  billiards has elongated
elliptical arcs. In Fig. 3(a-d) we show Poincar\' e sections  for
$\delta=0.19$  and  different values of $\gamma<0.982$. Similar results
for $\delta=0.60$  and $\gamma\le 0.80$  are shown in Fig. 4(a-d). These
pictures reveal a  highly chaotic behavior.  There are no elliptic
islands, however, flights  of points typical for neutral orbits can be
discerned. This is remarkably different from the corresponding results
for the elliptical stadium billiards\cite{LPESB}, where in the same
parameter region there were many fixed points and elliptic islands due
to stable pantographic and other orbits.

\begin{figure}
\end{figure}

\begin{figure}
\caption{\label{fig:Fig3}Poincar\' e plots for $\delta=0.19$ and  various
$\gamma$.                              }
\end{figure}

\begin{figure}
\end{figure}

\begin{figure}
\caption{\label{fig:Fig4}Poincar\' e plots for $\delta=0.60$ and  various
$\gamma$.                                   }
\end{figure}

\subsubsection{The bow-tie orbit}

We investigate the existence and stability of the bow-tie orbit (the
lowest pantographic orbit), shown in Fig. 5(a). This orbit exists if the
coordinates $x$ and $y$ of the impact point  and the derivative $y'$ of
the boundary at this point satisfy the equation (7), which now reads

\begin{equation}  
2yy'+x(1-y'^2)=0 
\end{equation}

\noindent giving as solution the coordinates of the
point of impact

\begin{equation} 
x=\sqrt{\frac{1-\delta^2-2\gamma^2}{1-\delta^2-\gamma^2}} 
\end{equation}

\noindent and 

\begin{equation}
y=\frac{\gamma^2}{\sqrt{(1-\delta^2)(1-\delta^2-\gamma^2)}}
\end{equation}

\noindent The condition $\delta<x<1$ that this point should lie on the
elliptical part of the boundary leads to the requirement 

\begin{eqnarray}
\nonumber\\
\delta<\sqrt{\frac{2-\gamma^2-\sqrt{\gamma^4+4\gamma^2}}{2}}; &&
\gamma<\frac{1-\delta^2}{\sqrt{2-\delta^2}}
\end{eqnarray}

\noindent This limit is shown in Fig. 2 and is denoted with the letter
a. If we denote the points with positive $x$ by 1 and the points on the
negative side by -1, the deviation matrix can be calculated as 

\begin{equation} 
M=(M_{11}M_{1-1})^2 
\end{equation} 

\noindent The corresponding angle $\alpha$ needed in the matrix (9) is  given by

\begin{equation} 
\sin\alpha=\sqrt\frac{1-\delta^2}{2(1-\delta^2-\gamma^2)}
\end{equation} 

\noindent The chords are
 
\begin{equation}  
\rho\equiv\rho_{1,1}=2y=\frac{2\gamma^2}
{\sqrt{(1-\delta^2)(1-\delta^2-\gamma^2)}}
\end{equation} 

\noindent and 

\begin{equation}  
\rho'\equiv\rho_{1,-1}=2\sqrt{x^2+y^2}=2
\sqrt{\frac{1-\delta^2-\gamma^2}{1-\delta^2}}
\end{equation}

\noindent  The curvature radius at the impact point is  obtained by
substituting (11) into (4) and reads 

\begin{equation}  
R=\frac{2\gamma^2\sqrt{2}}{1-\delta^2}
\end{equation}

\noindent If we
define 

\begin{equation}  
\Phi=\frac{\rho}{R\sin\alpha}\frac{\rho'}{R\sin\alpha}-(\frac{\rho}
{R\sin\alpha}+
\frac{\rho'}{R\sin\alpha})
\end{equation}

\noindent the trace of the deviation matrix is 

\begin{equation}  
{\rm Tr}M=2\left [ 2(2\Phi+1)^2-1  \right ]  
\end{equation} 

\noindent The left-hand side of the stability condition (8) is valid
automatically, but the right-hand side is fulfilled only if

\begin{equation}
-1<\Phi<0
\end{equation}

\noindent By substituting (15), (16), (17) and (18) into (19), one
obtains  $\Phi=-1$  for all allowed cases. The conclusion is that the
bow-tie orbit is neutral for all parameter values satisfying the
existence condition. 

\begin{figure}
\caption{\label{fig:Fig5}Typical lowest order periodic orbits in 
TEB billiards:  (a) bow-tie orbit (neutral); (b) rectangular orbit
(neutral);  (c) horizontal two-bounce orbit (stable);  (d) tilted
two-bounce orbit (neutral); (e) diamond orbit (stable); (f) multidiamond
orbit with $n=2$ (stable); (g) hour-glass orbit (neutral); (h) 8-shaped
orbit (stable). }
\end{figure}

\subsubsection{The rectangular orbit}

Further we investigate  properties of the rectangular orbit shown in
Fig. 5(b). According  to (7), this orbit exists if the  derivative on
the boundary is $y'=-1$. Corresponding solutions for the impact point
are

\begin{equation} 
x=\sqrt{\frac{1-\delta^2}{1-\delta^2+\gamma^2}}
\end{equation}

\noindent and 

\begin{equation} 
y=\frac{\gamma^2}{\sqrt{(1-\delta^2)(1-\delta^2+\gamma^2)}} 
\end{equation}

\noindent The condition $\delta<x<1$ leads to the requirement 

\begin{eqnarray}
\nonumber\\
\delta<\frac{\sqrt{\gamma^2+4}-\gamma}{2}; &&
\gamma<\frac{1-\delta^2}{\delta}
\end{eqnarray}

\noindent This limit is shown in Fig. 2 denoted by letter f. Stability
is calculated with equation (9) and the matrix (14), where the angle
$\alpha$ is given by $\sin\alpha=1/\sqrt{2}$. The chords are
$\rho\equiv\rho_{1,1}=2x$ and  $\rho'\equiv\rho_{1,-1}=2y$ and the
curvature radius is 

\begin{equation}
   R=\frac{2\sqrt{2}\gamma^2}{1-\delta^2}\left[
\frac{1-\delta^2}{1-\delta^2+\gamma^2}\right]^{3/2} 
\end{equation}

\noindent Again, the trace is given by (20), and for this case one
obtains  $\Phi=0 $. The conclusion is that also this orbit is neutral 
for all shapes, both flattened and elongated, allowed by (24).

The elongated truncated elliptical billiards were discussed in
\cite{DMTEB}. Their boundary shapes were described by means of two
parameters $h$ and $a$, related to our parameters $\delta$ and $\gamma$
as follows:

\begin{eqnarray}
\nonumber\\
h=\sqrt{1-\delta^2}; &&
a=\frac{\sqrt{1-\delta^2}}{\gamma} 
\end{eqnarray}

\noindent In \cite{DMTEB} the billiards with $a>1$ and $h<1$ have been
analysed and  the hyperbolic behavior has been identified in the region
$h<\rm{min}(1/a,1/\sqrt{2})$. In our parameters, this corresponds to
the  quasi-triangular region in the parameter space, denoted by A and B
in Fig. 2, delimited by curves $\delta=\sqrt{1-\gamma}$ (denoted in Fig.
2 by letter e) and $\delta=\sqrt{1-\gamma^2}$ and by the straight line
$\delta=1/\sqrt{2}$.

The region  $h<1/\sqrt{1+a^2}$ is rigorously proved to be
ergodic\cite{DMTEB}.   Written with our parameters, it obeys the
conditions

\begin{equation}
\frac{\sqrt{\gamma^4+4}-\gamma}{2}<\delta<\sqrt{1-\gamma^2}; 
\end{equation}

\begin{equation}
\frac{1-\delta^2}{\delta}<\gamma<\sqrt{1-\delta^2}
\end{equation}

\noindent The corresponding part of the parameter space in Fig. 2 is the
one denoted by A. The comparison with our results shows that the limit
(27) or (28) is identical to the limit (24) in the parameter space,
where the parabolic rectangular orbits emerge.

\subsection{Billiards with $\delta>\sqrt{1-\gamma^2}$}

This part of the parameter space, with flattened elliptical arcs, had
not been investigated previously. In Fig. 3(e-h) we show Poincar\' e
sections  for $\delta=0.19$  and  different values of $\gamma>0.98$. In
this parameter region dynamics is following the KAM scenario. Similar
behavior is noticed for values $\delta=0.60$  and  $\gamma>0.89$ (Fig.
4(e-h)). Elliptic islands corresponding to the horizontal two-bounce and
some other orbits are visible, similarly to the  corresponding results
for the elliptical stadium billiards\cite{LPESB}. We investigate the
existence and stability criteria for these orbits. 

\subsubsection{Horizontal diametral two-bounce orbits}

The horizontal two-bounce orbit (Fig. 5(c)) obviously exists for all
combinations of $\delta$ and $\gamma$, but according to
\cite{Berry,LPESB} the stability  condition 

\begin{equation}
\frac{\rho}{2R}<1 
\end{equation}

\noindent takes the form

\begin{eqnarray}
\nonumber\\
\delta>\sqrt{1-\gamma^2} ; &&
\gamma>\sqrt{1-\delta^2} 
\end{eqnarray}

\noindent so that bifurcations giving birth to stable diametral orbits
appear at the values $\delta=\sqrt{1-\gamma^2}$, corresponding to
circular arcs. In the Poincar\' e diagrams this orbit and the
surrounding quasiperiodic orbits are visible as two large bands near
$v_x=\pm 1$.

\subsubsection{Tilted diametral two-bounce orbits}

According to (7) a tilted two-bounce orbit (Fig. 5(d)) exists at the
point T$(x,y)$ on  the billiard boundary with derivative $y'$ if 

\begin{equation}
yy'+x=0 
\end{equation}

\noindent This is realized for any $\delta<x<1$ provided that

\begin{equation}
\gamma^2+\delta^2=1
\end{equation}

\noindent thus only for the truncated circle. Since in this case the
chord in (29) is $\rho=2$ and the radius is $R=1$, these orbits are
neutral. 

\subsubsection{Diamond orbit}

The diamond orbit of period four, shown in Fig. 5(e), exists for any
parameter choice. It has two bouncing points at the ends of the
horizontal semiaxis, and the other two on the flat parts on the
boundary.  To assess its stability, one should calculate the stability
matrix $M=(M_{01}M_{10})^2$. The angles contained in the matrix are
given as 

\begin{eqnarray}
\nonumber\\
\sin\alpha_0=\frac{\gamma}{\rho}; &&
\sin\alpha_1=\frac{1}{\rho}
\end{eqnarray}

where 

\begin{equation}
\rho=\sqrt{1+\gamma^2}
\end{equation}

\noindent The curvature radius at $x=1$ is given by (6). This leads to
the trace

\begin{equation} 
{\rm Tr}M=2\left [ 2\left ( \frac{2\rho^2}{R}-1 \right )^2-1 \right ] 
\end{equation}

\noindent and to the condition $\rho^2<R$ or
 $1+\gamma^2<\gamma^2/(1-\delta^2)$, thus the stable diamond orbit
 appears when

\begin{eqnarray}
\nonumber\\
\delta>\frac{1}{\sqrt{1+\gamma^2}}; &&
\gamma>\sqrt{\frac{1}{\delta^2}-1} 
\end{eqnarray}

\noindent This limit is shown in Fig. 2 as the line denoted by letter c.

\subsubsection{Multidiamond orbits}

The multidiamond orbit of order $n$ is the orbit of period $2+2n$, 
 which has two bouncing points at the ends of the horizontal axis and
 $2n$ bouncing points on the flat parts of the boundary (Fig. 4 (f)).
 Such orbit exists if

\begin{equation}
\delta>1-\frac{1}{n}
\end{equation} 
 
\noindent As explained for a similar case in \cite{LPESB}, the chord  $\rho$ 
in (35) should be replaced by

\begin{equation} 
 L=n\rho_n 
\end{equation}

\noindent where, for the truncated elliptical billiard, 

\begin{equation}
\rho_n=\sqrt{\frac{1}{n^2}+\gamma^2}
\end{equation}    

\noindent The trace of the stability matrix is then 

\begin{equation} 
{\rm Tr}M=2\left [ 2\left ( \frac{2\rho_n^2n^2}{R}-1 \right )^2-1 \right
]  
\end{equation}

\noindent with $R$ given by (6). The resulting condition for the
stability of the multidiamond orbit is  

\begin{eqnarray}
\nonumber\\
\delta>\sqrt{1-\frac{\gamma^2}{1+\gamma^2n^2}} ; &&
\gamma>\sqrt{\frac{1-\delta^2}{1-n^2(1-\delta^)}}
\end{eqnarray}

\noindent The limiting curves (41) are plotted in Fig. 2. The line with
$n=2$ is denoted by letter d,  and  above it there are several lines for
$n>2$. For $\gamma\to\infty$ the minimal values of $\delta$ above which
the multidiamond orbits appear are

\begin{equation} 
\lim \limits_{\gamma \to\infty}\delta=\sqrt{1-\frac{1}{n^2}} 
\end{equation}

\begin{figure}
\end{figure}

\begin{figure}
\caption{\label{fig:Fig6}Poincar\' e plots for a set of different 
values $\delta=\gamma$, showing the appearance of successive
multidiamond orbits of higher order. Two symmetrical triangle-shaped
islands for $\delta=\gamma=0.90$ originate from the "8-shaped" orbit. }
\end{figure}

\noindent The emergence of multidiamond orbits can be followed by
observing the Poincar\' e sections for a set of shapes with
$\delta=\gamma$ (Fig. 6). The values of this parameter for which an
orbit of new $n$ appears are given as intersections of the straight line
$\delta=\gamma$  with curves (41), and obey the equation

\begin{equation} 
n^2\delta^4-(n^2-2)\delta^2-1=0 
\end{equation}

\noindent For the diamond orbit ($n=1$) this equation reads

\begin{equation} 
\delta^4+\delta^2-1=0 
\end{equation}

\noindent and the orbit appears for 
$$\delta=\gamma=\sqrt{(\sqrt{5}-1)/2}=0.78615.$$
For
the same type of boundary the stable two-bounce orbit appeared at
$$\delta=\gamma=1/\sqrt{2}=0.707107.$$

\subsubsection{The hour-glass orbit}

The hour-glass orbit (Fig. 5(g)) looks like the bow-tie orbit rotated by
$\pi/2$. It  exists if the coordinates $x$ and $y$ of the impact point 
and the derivative $y'$ of the boundary at this point satisfy the
equation

\begin{equation}  
2xy'+yy'^2-y=0 
\end{equation}

\noindent giving as solution the coordinates of the impact point 

\begin{equation} 
x=\sqrt{\frac{1-\delta^2}{\delta^2+\gamma^2-1}}
\end{equation}

\noindent and

\begin{equation} 
y=\gamma\sqrt{\frac{2\delta^2-2+\gamma^2}{(1-\delta^2)(\delta^2+
\gamma^2-1)}}
\end{equation}

\noindent The condition $\delta<x<1$ that this point should lie on the
elliptical part of the boundary leads to the requirement 

\begin{equation}
\nonumber\\
\sqrt{1-\frac{\gamma^2}{2}}<\delta<
\sqrt{\frac{-\gamma^2+\sqrt{\gamma^2+4}}{2}}
\end{equation}

\begin{eqnarray}
\sqrt{2(1-\delta^2)}<\gamma<\frac{\sqrt{1-\delta^4}}{2} 
\end{eqnarray}

\noindent These limits define the region shown in Fig. 2 denoted by letter b.
 
If we denote the points with positive $y$ by 1 and the points on the
negative side by -1, the deviation matrix can be calculated from (14).
The angle $\alpha$ needed in the calculation is given as

\begin{equation} 
\sin\alpha=\sqrt{\frac{\gamma^2}{2(\delta^2+\gamma^2-1)}}
\end{equation} 

\noindent The curvature radius at this point is given as 

\begin{equation}
R=\sqrt{\frac{8(1-\delta^2)}{\gamma^2}}
\end{equation}

The chords are 
\begin{equation}  
\rho\equiv\rho_{1,1}=2x=2\sqrt{\frac{1-\delta^2}{\delta^2+\gamma^2-1}}
\end{equation} 

\noindent and 

\begin{equation}  
\rho'\equiv\rho_{1,-1}=2\sqrt{x^2+y^2}
=2\sqrt{\frac{\delta^2+\gamma^2-1}{1-\delta^2}}
\end{equation}

\noindent If we define $\Phi$ as in (19), the trace of the deviation
matrix  is again given by (20) and the orbit is stable if $-1<\Phi<0$.

When we substitute the calculated values of $R$, $\rho$, $\rho'$ and
$\sin\alpha$ into (19),  we obtain $\Phi=-1$ for all  allowed shapes and
conclude that the hour-glass orbit is neutral. This means that in the
truncated elliptical billiards (TEB) there  is no stable hour-glass
orbit, at variance with the elliptical stadium billiard (ESB), where such
an orbit having interesting properties was stable in a large fraction of
the parameter space\cite{LPESB}.  Besides the diamond and multidiamond
orbits, in Fig. 6 one discerns the presence of another, "8-shaped",
stable orbit, shown in Fig. 5(h).

\section{The box-counting numerical analysis of the degree of chaoticity 
in the full parameter space}

In this section we return to the question of limits within which the
truncated elliptical billiard is fully chaotic. Here we test these limits
numerically, with the help of the box-counting
method\cite{LPESB,DullRW,RoBC}. We calculate the Poincar\' e sections  for
a chosen pair of shape parameters, starting with $n_1$ randomly chosen
sets of initial conditions and iterating each orbit for $n_2$
intersections with the x-axis, thus obtaining $n_1\times n_2$ points in
the Poincar\' e diagram. Then we divide the first quadrant of the phase
plane into a grid of $n\times n$ squares (boxes), count the number of
boxes which have points in them and calculate the ratio of this number to
the total number of boxes. The obtained ratio is denoted by $q_{\rm
class}$. In this way also certain points belonging to invariant curves
within the regular islands are included. But since our main aim is to
examine the onset of full chaos, this method gives satisfactory results,
providing that the appropriate values of $n_1$, $n_2$ and $n$ are used.
Detailed testing has shown that reliable results are obtained for values
$n_1=100$, $n_2=5000$ and $n=100$ used in our present calculation\cite{Simic}.

\begin{figure}
\caption{\label{fig:Fig7}Diagram showing the degree of chaoticity
$q_{\rm class}$  of the truncated elliptical billiards (TEB), in
dependence on the shape parameters, with $0<\gamma<2.2$ and $0\le
\delta\le 1$. Black points denote shapes with $q_{\rm class}=1.00$, red
with $0.90\le q_{\rm class}<1.00$, green with $0.80\le q_{\rm
class}<0.90$, yellow with $0.70\le q_{\rm class}<0.80$, blue with 
$0.60\le q_{\rm class}<0.70$ and grey with $0.00\le q_{\rm class}<0.60$.}
\end{figure}

\begin{figure}
\caption{\label{fig:Fig8}Diagram showing the degree of chaoticity
$q_{\rm class}$ of the  elliptical stadium billiards (ESB), in dependence
on the shape parameters, with $0<\gamma<2.2$ and $0\le \delta\le 1$.
Black points denote shapes with $q_{\rm class}=1.00$,  red with $0.90\le
q_{\rm class}<1.00$, green with $0.80\le q_{\rm class}<0.90$, yellow with
$0.70\le q_{\rm class}<0.80$, blue with  $0.60\le q_{\rm class}<0.70$ and
grey with $0.00\le q_{\rm class}<0.60$.}
\end{figure}

In Fig. 7 we plot in the $\delta-\gamma$ plane the points  representing
the pairs of shape parameters. Points are plotted in different colors,
depending on the corresponding value of $q_{\rm class}$. The full chaos,
corresponding to $q_{\rm class}=1.00$, is depicted by black points.
Colored points denote  shapes within intervals between 0 and 0.99. This 
diagram confirms that for the truncated elliptical billiards (TEB), in
the region below  the onset of the  stable two-bounce horizontal orbit,
dynamics is practically completely chaotic. This is in strong contrast
with the behavior of the elliptical stadium billiard for which the
similar diagram is  shown in Fig. 8. For the ESB billiards the region of
chaos was strictly bounded also  from the lower side and determined by
emergence of stable pantographic orbits.

To examine the possible mechanism for this difference, we assume that the
TEB and the  ESB billiards are two extreme cases and search for a
possible transition between them.

\section{Generalized  truncated stadium-like elliptical billiards}

In this section we propose a new large class of stadium-like billiards
which we call generalized truncated elliptical  stadium-like billiards
(GTESB). Such a billiard depends on three shape parameters $\delta$,
$\gamma$ and $\kappa$. The allowed values of the shape parameters are 

\begin{eqnarray}
\nonumber\\
0\le\delta\le1; &
0<\gamma<\infty ;&
-1\le\kappa \le 1 
\end{eqnarray}

\begin{figure}  
\caption{\label{fig:Fig9}Construction and parameters of the
 generalized elliptical truncated stadium-like billiards (GTESB)}
\end{figure} 

\noindent For the limiting values of $\kappa$  we obtain the two billiard
families considered before: for $\kappa=-1$ GTESB reduces to the
elliptical stadium billiard (ESB), and for $\kappa=1$ GTESB becomes the
truncated elliptical billiard (TEB).

The new GTESB billiard boundary is obtained by adding elliptical arcs
symmetrically at the two opposite ends of a rectangle with sides
$2\delta$ and $2\gamma$. Elliptical arcs are cut out from the two
identical but generally detached ellipses by two horizontal straight
lines at $y=\pm\gamma$ (Fig. 9).  The two ellipses have centers at the
points

\begin{equation}
X_{\pm}=\pm\delta \left( \frac{1-\kappa}{2}\right)
\end{equation}

\noindent The distance between the two centers is $D=\delta(1-\kappa)$.
 The horizontal and vertical semiaxis are given, respectively, as 

\begin{equation}
A_{\rm x}=1-\delta\left (\frac{1-\kappa}{2}\right )
\end{equation}

\noindent and

\begin{equation}
A_{\rm
y}=\frac{\gamma}{\sqrt{(1-\delta)(1+\kappa\delta)}}
\left[1-\delta\left(\frac{1
-\kappa}{2}\right)\right]
\end{equation}

\noindent and the equation of the two ellipses reads

\begin{equation}
\left(\frac{x-X_{\pm}}{A_{\rm x}}\right)^2+
\left(\frac{y}{A_{\rm y}}\right)^2=1
\end{equation}

\noindent The horizontal diameter of the billiard is 2. For $\kappa=-1$
 and $\gamma=1-\delta$ the GTESB  becomes the Bunimovich stadium
 billiard.

\begin{figure}
\end{figure}

\begin{figure}
\caption{\label{fig:Fig10} Poincar\' e plots for the  generalized
truncated elliptical stadium-like billiards  (GTESB), for $\delta=0.2$
and $\gamma=0.6$ with different values of $\kappa$: (a) $\kappa=-1.0$; (b)
$\kappa=-0.6$; (c) $\kappa=-0.3$; (d) $\kappa=0$;(e) $\kappa=0.3$;  (f)
$\kappa=0.6$; (g) $\kappa=0.90$; (h) $\kappa=1.0$.}
\end{figure}

In Fig. 10 the Poincar\' e sections are shown for $\delta=0.2$ and
$\gamma=0.6$ with $\kappa$ assuming different values between -1 and 1.
The four islands typical for the bow-tie orbit are present for all
$\kappa$ except for $\kappa=1$ (TEB), where this orbit becomes neutral
and the island reduces to a caracteristical flight of points. The limits
separating chaotic from mixed behavior are determined by the onset of the
stable horizontal 2-bounce orbit and are given by (29).  Since the
curvature radius at $|x|=1$ is 

\begin{equation}
R_1=\frac{\gamma^2[2-\delta(1-\kappa)]}{2(1-\delta)(1+\delta\kappa)}
\end{equation}

\noindent the upper limit of chaos is determined by the condition

\begin{equation}
\gamma>\sqrt{\frac{2(1-\delta)(1+\kappa\delta)}{2-\delta(1-\kappa)}}
\end{equation}

\begin{figure}
\caption{\label{fig:Fig11}Dependence of $q_{\rm class}$ on $\gamma$ for
 a set of shapes with $\delta=\gamma$ and the choice of $\kappa$ as in Fig.10. }
\end{figure}

\noindent In Fig. 11 the chaotic fraction $q_{\rm class}$ is shown for
the special case $\delta=\gamma$ for different $\kappa$, in dependence on
$\gamma$. It is noticed that in the case $\kappa=-1$ (ESB) there is a
narrow, strictly limited region of full chaos, outside of which the
values of $q_{\rm class}$ are low. For $\kappa=1$ (TEB) the fully chaotic
region is much larger and  extends practically over all values $\delta$
and $\gamma$ below the chaotic limit.  Between these two limits, the
regions of full chaos ($q_{\rm class}=1$) are shorter and limited, but
there are many shapes with chaotic parameter close to 1 (between 0.90 and
0.99). This corresponds to a selection of narrow islands in the Poincar\'
e plots, as seen in Fig. 10.

\section{Discussion and conclusions}

In conclusion, our investigation of the elliptical stadium-like billiards
has revealed a rich variety of integrable, mixed and chaotic behavior,
which is connected with the character of the two elliptical arcs and with
their mutual position. This strong dependence  on parameters $\delta$ and
$\gamma$ is confirmed for the truncated elliptical billiards, but is even
more enhanced when a third shape parameter $\kappa$ is added. Analysis
shows, however, that among all considered shapes the truncated elliptical
billiard (TEB),  created by cutting a single ellipse with two parallel
straight lines, has exceptional properties, notably that it is chaotic
practically in the whole region of elongated elliptical arcs.  Notable is
the presence of many neutral orbits in this region, consistent with the 
fact that these orbits actually can be identified as orbits in an
ellipse. For the flattened arcs, the stable islands due to the two-bounce
horizontal orbit and to the diamond and multidiamond  orbits occupy an
important part of the phase plane. 

Our investigations can be useful for the experimental application of
billiards in the laser technology, where properties and directional
intensities of the optical microresonators depend strongly on the
boundary  shape. They can also be applied in designing the semiconducting
optical devices and in the technology of microwave and acoustic resonant
cavities. With this purpose in mind, we propose further analysis  of the
stadium-like billiards with elliptical arcs and an extension of the
present investigation to different types of open billiards.

\section{Acknowledgments}

Authors are thankful to A. B\" acker, M. Lebental, N. Pavin, T. Prosen,
M. Robnik and T. Tanaka for useful discussions and comments and to V.
Danani\' c  and D. Radi\' c for help with numerical  methods and
computation.

\end{document}